\documentclass[prb,twocolumn,showpacs,preprintnumbers,amsmath,amssymb]{revtex4}
\usepackage{graphicx}
\usepackage{dcolumn}
\usepackage{epsfig}

\begin{document}

\title{Evidence for a quantum phase transition in electron-doped Pr$_{2-x}$Ce$_{x}$CuO$_{4-\delta}$ from thermopower measurements}

\author{Pengcheng Li$^1$}
 \email{pcli@physics.umd.edu}
\author{K. Behnia$^2$}
\author{R. L. Greene$^1$}
\affiliation{$^1$Center for Superconductivity Research and
Department of Physics, University of Maryland, College Park,
Maryland 20742-4111, USA \\$^2$Laboratoire de Physique Quantique
(CNRS), ESPCI 10 Rue Vauquelin, F-75005 Paris, France} \

\date{\today}

\begin{abstract}
The evidence for a quantum phase transition under the
superconducting dome in the high-$T_c$ cuprates has been
controversial. We report low temperature normal state
thermopower(S) measurements in electron-doped
Pr$_{2-x}$Ce$_{x}$CuO$_{4-\delta}$ as a function of doping
(\emph{x} from 0.11 to 0.19). We find that at 2 K both S and S/T
increase dramatically from \emph{x}=0.11 to 0.16 and then saturate
in the overdoped region. This behavior has a remarkable similarity
to previous Hall effect results in
Pr$_{2-x}$Ce$_{x}$CuO$_{4-\delta}$. Our results are further
evidence for an antiferromagnetic to paramagnetic quantum phase
transition in electron-doped cuprates near \emph{x}=0.16.
\end{abstract}

\maketitle

The existence of a quantum phase transition at a doping under the
superconducting dome in high-$T_c$ superconductors is still
controversial. Evidence for a quantum critical point has been
given for hole-doped cuprates\cite{Tallon, Loram, Ando} but the
T=0 normal state is difficult to access because of the large
critical field(H$_{c2}$). Electron-doped cuprates have a
relatively low H$_{c2}$ and several studies have suggested that a
quantum phase transition exists in those cuprates. Electrical
transport\cite{Yoram} on electron-doped
Pr$_{2-x}$Ce$_{x}$CuO$_{4-\delta}$(PCCO) shows a dramatic change
of Hall coefficient around doping $x_c$=0.16, which indicates a
Fermi surface rearrangement at this critical doping. Optical
conductivity experiments\cite{Zimmers} revealed that a
density-wave-like gap exists at finite temperatures below the
critical doping $x_c$ and vanishes when $x\geq x_c$. Neutron
scattering experiments\cite{Kang} on
Nd$_{2-x}$Ce$_{x}$CuO$_{4-\delta}$(NCCO) found antiferromagnetism
as the ground state below the critical doping while no long range
magnetic order was observed above $x_c$. Other suggestive
evidence\cite{Fournier} comes from the observation of a low
temperature normal state insulator to metal crossover as a
function of doping, and the disappearance of negative spin
magnetoresistance at a critical doping\cite{Yoramupturn}. All
these experiments strongly suggest that an antiferromagnetic(AFM)
to paramagnetic quantum phase transition(QPT) occurs under the
superconducting dome in the electron-doped cuprates.

The quantum phase transition in electron-doped cuprates is
believed to be associated with a spin density wave(SDW) induced
Fermi surface reconstruction\cite{Lin,Zimmers}. Angle resolved
photoemission spectroscopy(ARPES) experiments\cite{Armitage} on
NCCO reveal a small electron-like pocket at$(\pi, 0)$ in the
underdoped region and both electron- and hole-like Fermi pockets
near optimal doping. This interesting feature is thought to arise
as a result of the SDW instability that fractures the conduction
band into two different parts\cite{Lin}. If one continues to
increase the doping(above $x_c$), the weakening of the spin
density wave leads to a large hole-like Fermi pocket centered at
$(\pi, \pi)$ in the overdoped region\cite{Lin,Matsui}.

Nevertheless, the presence of a quantum critical point(QCP) under
the superconducting dome in electron-doped cuprates is still quite
controversial\cite{Greven}. Other experimental probes of the
critical region are needed. In this paper, we present a systematic
study of the magnetic field driven normal state thermopower on
PCCO films. We find a doping dependence similar to that seen in
the low temperature normal state Hall effect
measurements\cite{Yoram}. From a simple free electron model
comparison of these two quantities, we find a strikingly similar
behavior of the effective number of carriers. This strongly
suggests that a quantum phase transition takes place near x=0.16
in PCCO.

High quality PCCO films with thickness about 3000\AA\ were
fabricated by pulsed laser deposition on SrTiO$_3$ substrates
(10$\times$5 mm$^2$). Detailed information can be found in our
previous papers\cite{Peng,Maiser}. The films were characterized by
AC susceptibility, resistivity measurements and Rutherford Back
Scattering(RBS).

High resolution thermopower is measured using a steady state
method by switching the temperature gradient to cancel the Nernst
effect and other possible background contributions. The sample is
mounted between two thermally insulated copper blocks. The
temperature gradient is built up by applying power to heaters on
each block and the gradient direction is switched by turning on or
off the heaters. The temperature gradient is monitored by two
Lakeshore Cernox bare chip thermometers. Thermopower data is taken
when the gradient is stable and averaged for many times to reduce
the systematic error. The voltage leads are phosphor bronze, which
has a small thermopower even at high field\cite{Wangyy}. The
thermopower contribution from the wire is calibrated against
YBa$_2$Cu$_3$O$_7$(T$_{c}$=92 K) for T$<$90 K and Pb film for
T$>$90 K, and is subtracted out to get the absolute thermopower of
the PCCO sample.

We measured the zero field and in field resistivity of all the
doped PCCO films. The results are similar to our previous
report\cite{Yoram}. A 9 T magnetic field(H$\parallel$c) is enough
to suppress the superconductivity for all the dopings. This
enables us to investigate the low temperature normal state
properties in PCCO. A low temperature resistivity upturn is seen
for doping below \emph{x}=0.16, which suggests a possible
insulator to metal crossover as a function of
doping\cite{Fournier}.

\begin{figure}
\centerline{\epsfig{file=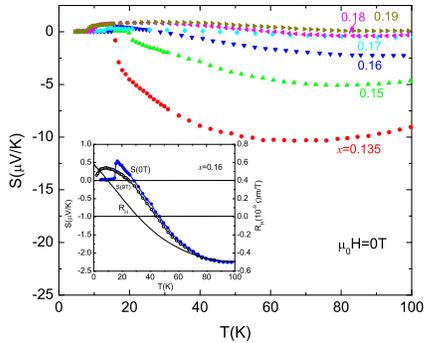,clip=,silent=,width=2.5in}}
\caption{(Color online) Thermopower S versus temperature(T$<$100
K) at zero field for all the superconducting PCCO films. Inset is
the thermopower S of \emph{x}=0.16 film at zero field(solid blue
circle) and $\mu_0$H=9 T(open circle) as a function of
temperature. Solid line is the temperature dependence of the Hall
coefficient for the same film.} \label{fig1}
\end{figure}

Thermopower is measured on the PCCO films doped from \emph{x}=0.11
to 0.19. In zero field, a sharp superconducting transition is
clearly seen in the thermopower. In the inset of Fig.~\ref{fig1},
we show the thermopower S of \emph{x}=0.16(T$_c$=16.5 K) as a
function of temperature. Our high resolution thermopower setup
enables us to observe small changes of signal. When the sample
goes to the superconducting state, S=0, a small change
$\triangle$S=0.5 $\mu$V/K is easily detectable, which indicates a
better sensitivity than our previous one-heater-two-thermometer
setup\cite{Budhani}. We also show the Hall coefficient R$_H$ as a
function of temperature for the same film in the graph. A sign
change of both S and R$_H$ is observed at the same temperature.

In the main panel of Fig.~\ref{fig1}, we show the zero field
thermopower for all the superconducting films. A clear
superconducting transition is seen in these films. The normal
state S(T$>$T$_c$) is negative in the underdoped region. It
becomes positive in the overdoped region at low temperature(to be
shown later). The magnitude of S in the underdoped region is large
as expected for a system with less charge carrier density while it
is much smaller in the overdoped region. Previous zero field
thermopower measurements on NCCO crystals\cite{Wang} are
qualitatively similar to our data.
\begin{figure}
\centerline{\epsfig{file=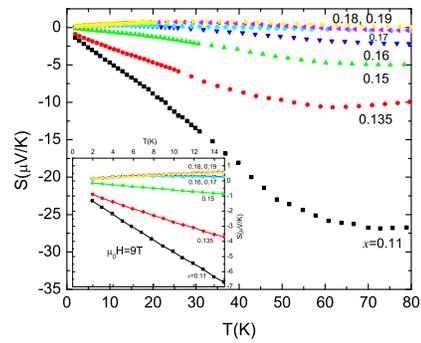,clip=,silent=,width=2.5in}}
\caption{(Color online)The normal state thermopower S($\mu_0$H=9
T$>\mu_0$H$_{c2}$) of all the doped films versus temperature.
Inset shows the low temperature (T$<$15 K) data.} \label{fig2}
\end{figure}

When a 9 T magnetic field is applied along the c-axis, the
superconducting films are driven to the normal state for
T$<$T$_c$. As seen from the inset of Fig.~\ref{fig1}, when the
superconductivity is destroyed, the normal state thermopower is
obtained. In Fig.~\ref{fig2}, we show the normal state thermopower
for all the films. The low temperature(T$<$15 K) normal state
thermopower  is shown in the inset. We showed in Fig.~\ref{fig1}
that for \emph{x}=0.16 the thermopower changes from negative to
positive for T$<$30 K, in good agreement with the Hall effect
measurements\cite{Yoram}. For the overdoped films \emph{x}=0.17
and 0.18, we observe similar behavior with a sign change occurring
below 45 K and 60 K respectively. However, the thermopower is
always positive for \emph{x}=0.19. Similar to the the Hall effect,
the thermopower for \emph{x}$\geq$0.16 is nearly same for T$<$10
K, as shown in the inset of Fig.~\ref{fig2}. The dramatic change
of the thermopower at low temperature from \emph{x}=0.15 to the
overdoped region suggests  a sudden Fermi surface rearrangement
around the critical doping \emph{x}=0.16.

\begin{figure}
\centerline{\epsfig{file=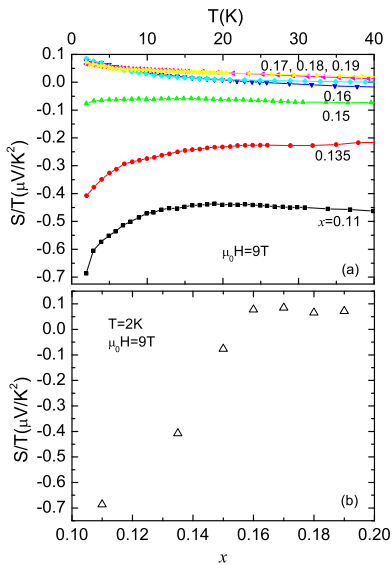,clip=,silent=,width=2.3in}}
\caption{(Color online)(a) S/T versus temperature (T$<$40 K and
$\mu_0$H=9 T) for all the films. (b) S/T(T=2 K and $\mu_0$H=9 T)
as a function of doping \emph{x}.} \label{fig3}
\end{figure}

\begin{figure}
\centerline{\epsfig{file=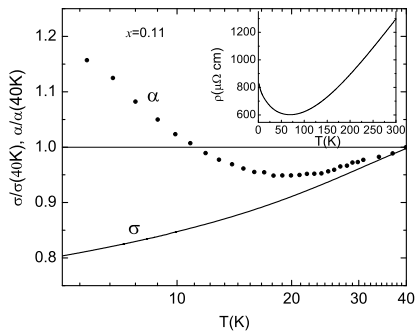,clip=,silent=,width=2.5in}}
\caption{Normalized $\alpha=S/T$ and $\sigma(T)$ for \emph{x}=0.11
versus temperature for T$\leq$40 K. Inset shows the temperature
dependence of in plane resistivity $\rho(T)$ for the same film.}
\label{fig4}
\end{figure}

In the Boltzmann picture, thermopower and electrical conductivity
are related through the expression\cite{Ashcroft}:
\begin{equation}\label{1}
S=\frac{-\pi^{2}k_{B}^{2}T}{3e}\frac{\partial{ln\sigma(\epsilon)}}{\partial{\epsilon}}|_{\epsilon=E_{F}}
\end{equation}
In the simple case of  a free electron gas,  this yields: $S/T=
\frac{-\pi^{2}k_{B}^{2}}{3e}\frac{N(\epsilon_F)}{n}$
(N($\epsilon_F$) is the density of states at the Fermi energy and
$n$ is the total number of charge carriers). However, in real
metals, the energy-dependence of the scattering time at the Fermi
level,
$(\frac{\partial\ln\tau(\epsilon)}{\partial\epsilon})_{\epsilon=\epsilon_{f}}$,
also affects the thermopower. In the zero-temperature limit, it
has been shown that this term also becomes proportional to
$\frac{N(\epsilon_F)}{n}$ when the impurity scattering
dominates\cite{Miyake}. In electron-doped cuprates, there is
strong evidence\cite{Yoram} for impurity scattering at low
temperatures. The residual resistivity is about 50 $\mu\Omega$-cm
for an optimally-doped film, which is quite large compared to
clean metals, and the temperature dependence of the resistivity
becomes almost constant below 20 K. This is all suggestive of
strong impurity scattering. The scattering most likely comes from
Ce and oxygen disorder and one would expect a similar disorder at
all dopings, although this is hidden by the anomalous (and
unexplained) resistivity upturn for the lower dopings. Therefore,
we expect that the thermopower is proportional to N(E$_F$)/n will
be a valid approximation for our electron-doped PCCO films. This
theory thus provides a solid theoretical basis for an experimental
observation: in a wide variety of correlated metals, there is an
experimental correlation between the magnitude of thermopower and
specific heat in the zero-temperature limit\cite{Behnia}.

Let us examine our data with this picture in mind.
Fig.~\ref{fig3}(a) presents S/T as a function of temperature below
40 K for all the doped films. As seen in the figure, there is a
dramatic difference between the underdoped and the overdoped
films. For underdoped, S/T displays a strong temperature
dependence below 20 K, which is reminiscent of the low temperature
upturn in resistivity and Hall effect\cite{Fournier,Yoram}.  One
possible explanation for this feature would be charge localization
\cite{Fournier3}. If all, or some of, the itinerant carriers
localize at very low temperatures, then the decrease in
conductivity is expected to be concomitant with an increase in the
entropy per itinerant carrier (which is the quantity roughly
measured by S/T). We find this to be qualitatively true as shown
in Fig.~\ref{fig4}, which displays S/T and conductivity for
\emph{x}=0.11 in a semilog plot. Below 10 K, both quantities are
linear functions of $\log$T. Note that for the resistivity, it has
been shown\cite{Fournier} that the logarithmic divergence
saturates below 1 K. Therefore, further thermopower measurements
below 2 K would be very useful.

In contrast to the underdoped films, the temperature dependence of
S/T in the overdoped region is weaker and there is clearly a
finite  S/T even at zero temperature. Taking the magnitude of S/T
at 2 K as our reference, we can examine the doping dependence of
the ratio $\frac{N(\epsilon_F)}{n}$ for itinerant carriers at this
temperature. Fig.~\ref{fig3}(b) presents the doping dependence of
S/T at 2 K. A strong doping dependence for $x\leq$0.16, a sharp
kink around \emph{x}=0.16 and a saturation in the overdoped region
are visible. The dramatic change of S/T at low temperatures from
the underdoped to overdoped regions is similar to the Hall
effect\cite{Yoram} at 0.35 K, in which a sharp kink was observed
around \emph{x}=0.16. Both S/T and R$_H$ change from negative in
the underdoped region to a saturated positive value above
\emph{x}=0.16.

\begin{figure}
\centerline{\epsfig{file=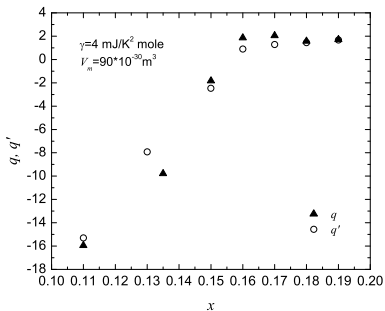,clip=,silent=,width=2.3in}}
\caption{Doping dependence of \emph{q}(2 K) and \emph{q'}(2 K) of
PCCO films(q and q' are defined by Eq. (2) and (3) in the text).}
\label{fig5}
\end{figure}

The similarity of the doping dependence of S/T and R$_H$ implies a
common physical origin. To explore the relation between S/T and
R$_H$, let us assume a simple free electron model, where
thermopower displays a very simple correlation with the electronic
specific heat, $C_{el}= \frac{\pi^{2}k_{B}^{2}T}{3}N(\epsilon_F)$.
Following the analysis of Ref.20, a dimensionless quantity
\begin{equation}\label{2}
q=\frac{S}{T}\frac{N_{Av}e}{\gamma}
\end{equation}
can be defined($N_{Av}$ is Avogadro's number and
$\gamma=C_{el}/T$), which is equal to $N_{Av}/n$. For a simple
metal, R$_H=V/ne$ ($V$ is the total volume). If we define
\begin{equation}\label{3}
q'=R_He/V_m
\end{equation}
where $V_m$ is unit cell volume, then $q'$ is also equal to
$N_{Av}/n$. By this simple argument, we can compare S and R$_H$
directly. Because we do not have data for $\gamma$ except at
optimal doping, we assume it does not change much with doping.
With the $\gamma$ value($4mJ/K^2mole$)\cite{Hamza} for
\emph{x}=0.15 and S/T and R$_H$ at 2 K, we can plot both $q$ and
$q'$ together, as shown in Fig.~\ref{fig5}. We find a remarkable
similarity in the doping dependence of these two dimensionless
quantities, both in trend and in magnitude. Note that no dramatic
changes in either $q$ or $q'$ are observed near \emph{x}=0.13,
where it is claimed that AFM long range order
vanishes\cite{Greven} from recent neutron scattering measurements.
We should mention that assuming a constant $\gamma$ as a function
of doping in our range of investigation (\emph{x}=0.11 to 0.19)
is, of course, subject to caution due to a lack of experimental
data. However, it has been found\cite{Hamza} that the specific
heat coefficient $\gamma$ is the same for an as-grown crystal and
a superconducting Pr$_{1.85}$Ce$_{0.15}$CuO$_4$ crystal. Neutron
scattering studies have shown that an as-grown \emph{x}=0.15
crystal is equivalent to an annealed Pr$_{1.88}$Ce$_{0.12}$CuO$_4$
crystal\cite{Greven2}. This strongly suggests that $\gamma$ will
not change much with Ce doping at least in the critical range
around optimal doping. Therefore, no significant change in the
doping dependence of $q$ due to this correction is expected.

We believe that the saturation of S/T in the overdoped region is a
result of the Fermi surface rearrangement due to the vanishing of
antiferromagnetism above a critical doping. To our knowledge,
there is no theoretical prediction for the doping dependence of
the thermopower in an antiferromagnetic quantum critical system.
Although the temperature dependence of thermopower near zero
temperature is given by Paul \emph{et al}.\cite{Paul} for such a
system near critical doping, we are not yet able to access the
very low temperature region(T$<$2 K) to test these predictions in
PCCO. Nevertheless, an amazing agreement between thermopower and
Hall effect measurements is shown in our simple free electron
model. This model is certainly oversimplified since there is
strong evidence for two types of carriers near optimal
doping\cite{Jiang,Fournier2, Gollnik}. But, much of this transport
data\cite{Jiang, Fournier2, Gollnik} implies that one type of
carrier dominates at low temperature. Thus a simple model may be
reasonable. However, to better understand this striking result a
more detailed theoretical analysis will be needed.

Interestingly, the number $q$ in overdoped PCCO is close to 1. It
was shown that when $q$ is close to unity, a Fermi liquid behavior
is found in many strongly correlated materials\cite{Behnia}. This
suggests that overdoped PCCO is more like a Fermi liquid metal
than underdoped PCCO. When x is above the critical doping
\emph{x}=0.16, $q$ and $q'$ are close to $1/(1-x)$, which suggests
that the hole-like Fermi surface is recovered in accordance with
local density approximation band calculations and the Luttinger
theorem.

In summary, we performed high resolution measurements to
investigate the low temperature normal state thermopower(S) of
electron-doped cuprates Pr$_{2-x}$Ce$_{x}$CuO$_{4-\delta}$(PCCO).
We find a strong correlation between S/T and the Hall coefficient
(R$_H$) at 2 K as a function of doping. Using a simple free
electron model, which relates thermopower to the electronic
specific heat, we conclude that our observations support the view
that a quantum phase transition occurs near \emph{x}=0.16 in the
PCCO system.

This work is supported by NSF Grant DMR-0352735. We thank Drs.
Andy Millis and Victor Yakovenko for fruitful discussions.

\end{document}